\documentclass[12pt,a4wide]{revtex4}
\usepackage{graphicx}
\usepackage{citesort}
\usepackage{epsf}
\usepackage{bm}

\newcommand{\la}{\left\langle}
\newcommand{\ra}{\right\rangle}
\newcommand{\dd}{\mbox{d}}

\begin{document}
\title{Acceleration and vortex filaments in turbulence}
\author{F. Toschi$^{1}$, L. Biferale$^{2}$, G.
  Boffetta$^{3}$, A.  Celani$^{4}$,\break B.~J. Devenish$^{2}$ and A.
  Lanotte$^{5}$} 
\affiliation{$^{1}${Istituto per le Applicazioni del Calcolo, CNR, Viale del Policlinico 137, I-00161 Roma, Italy}\\
$^{2}${Dipartimento di Fisica, Universit\`a degli Studi di
    Roma ``Tor Vergata'' and INFN, Via della Ricerca Scientifica 1,
    I-00133 Roma, Italy}\\
$^{3}${Dipartimento di Fisica
    Generale, Universit\`a degli Studi di Torino and INFN,\break Via
    Pietro Giuria 1, I-10125, Torino, Italy}\\
$^{4}${CNRS,
    INLN, 1361 Route des Lucioles, 06560 Valbonne, France}\\
$^{5}${CNR-ISAC, Str.  Prov. Lecce-Monteroni km. 1200,
    I-73100 Lecce, Italy}}


\begin{abstract}
  We report recent results from a high resolution numerical study of
  fluid particles transported by a fully developed turbulent flow.
  Single particle trajectories were followed for a time range spanning
  more than three decades, from less than a tenth of the Kolmogorov
  time-scale up to one large-eddy turnover time. We present some
  results concerning acceleration statistics and the statistics of
  trapping by vortex filaments.
\end{abstract}
\maketitle 
\newpage 

Lagrangian statistics of particles advected by a turbulent velocity
field, $\bm u(\bm x,t)$, are important both for their theoretical
implications \cite{K65} and for applications, such as the development
of phenomenological and stochastic models for turbulent mixing
\cite{pope}. Despite recent advances in experimental techniques for
measuring Lagrangian turbulent statistics
\cite{cornell,pinton,ott_mann,leveque}, direct numerical simulations
(DNS) still offer higher accuracy albeit at a slightly lower Reynolds
number \cite{yeung,BS02,IK02}. Here, we describe Lagrangian statistics
of velocity and acceleration in terms of the multifractal formalism.
At variance with other descriptions based on equilibrium statistics (see e.g.
\cite{beck,aringazin,arimitsu}, critically reviewed in \cite{GK03}),
this approach has the advantage of being founded on solid
phenomenological grounds. Hence, we propose a derivation of the
Lagrangian statistics directly from the Eulerian statistics.

We analyze Lagrangian data obtained from a recent Direct Numerical
Simulation (DNS) of forced homogeneous isotropic turbulence
\cite{biferale04,biferale04b} which was performed on $512^3$ and
$1024^3$ cubic lattices with Reynolds numbers up to $R_\lambda \sim
280$. The Navier-Stokes equations were integrated using fully
de-aliased pseudo-spectral methods for a total time $T \approx T_L$.
Two millions of Lagrangian particles (passive tracers) were injected
into the flow once a statistically stationary velocity field had been
obtained. The positions and velocities of the particles were stored at
a sampling rate of $0.07 \tau_\eta$. The velocity of the Lagrangian
particles was calculated using linear interpolation.  Acceleration was
calculated both as the derivative of the particle velocity and by
direct computation from all three forces acting on the particle (i.e.
pressure gradients, viscous forces and large scale forcing): the two
measurements were found to be in very good agreement.  Finally, the
flow was forced by keeping the total energy constant in each of the
first two wavenumber shells.  For more details on the simulation, see
\cite{biferale04,biferale04b}.

\section{Velocity and acceleration statistics}
Velocity statistics along a particle trajectory can be measured by
means of the Lagrangian structure functions, $S_p(\tau) = \langle
(\delta_\tau v)^p \rangle$ where $\delta_{\tau} v$ is the Lagrangian
increment of one component of the velocity field in a time lag
$\tau$. A simple way to link the Lagrangian velocity increment,
$\delta_{\tau} v$, to the Eulerian one, $\delta_r u$, is to consider
the velocity fluctuations along a particle trajectory as the
superposition of different contributions from eddies of all sizes.  In
a time-lag $\tau$ the contributions from eddies smaller than a given
scale, $r$, are uncorrelated, and we may write $\delta_{\tau} v \sim
\delta_r u$. Assuming that typical eddy turn over time $\tau$ at a
given spatial scale $r$ can be expressed as $\tau_r \sim r/\delta _r u
$, one obtains:
\begin{equation}
\label{eq:tau_l}
\delta_{\tau} v \sim \delta_r u 
\qquad \tau \sim  \frac{L^h_0}{v_0} r^{1-h},
\end{equation}
where $h$ is the local scaling exponent characterizing the Eulerian
fluctuation in the multifractal phenomenology \cite{frisch}. Also,
$L_0, v_0$ are the integral scale and the typical velocity,
respectively.  With respect to the the usual multifractal
phenomenology of fully developed turbulence, the presence of
a fluctuating eddy turn over time is the only extra additional
ingredient to take into account in the Lagrangian reference frame.

Using (\ref{eq:tau_l}), one can estimate the
Lagrangian velocity structure function:
\begin{equation}
\label{eq:sfl}
S_p(\tau) \sim \langle v_0^p \rangle \int_{h \in I} \dd h \left(
  \frac{\tau}{T_L} \right)^{\frac{hp + 3-D(h)}{1-h}},
\end{equation}
where the factor $(\tau/T_L)^{(3-D(h))/(1-h)}$ is the probability of
observing an exponent $h$ in a time-lag $\tau$, and $D(h)$ is the
dimension of the fractal set where the exponent $h$ is observed.  The
Lagrangian scaling exponents $\zeta_L(p)$ can be estimated by a saddle
point approximation, for $\tau \ll T_L$:
\begin{equation}
\zeta_L(p) = \inf_h \left( \frac{hp + 3-D(h)}{1-h} \right).
\label{eq:zeta_l}
\end{equation}
We would like to stress that for the $D(h)$ curve we have chosen that
of the Eulerian statistics. In other words, the prediction
(\ref{eq:zeta_l}) is free of any additional parameter once the
Eulerian statistics are assumed \cite{borgas93,BDM02,biferale04}.\\
%%%%%%%%%%%%%%%%%%%%%%%%%%%%%%%%%%%%%%%%%%%%%%%%%%%%%%%%%%%%%%%%
\begin{figure}[!t]
\begin{center}
  \includegraphics[draft=false,scale=1.0]{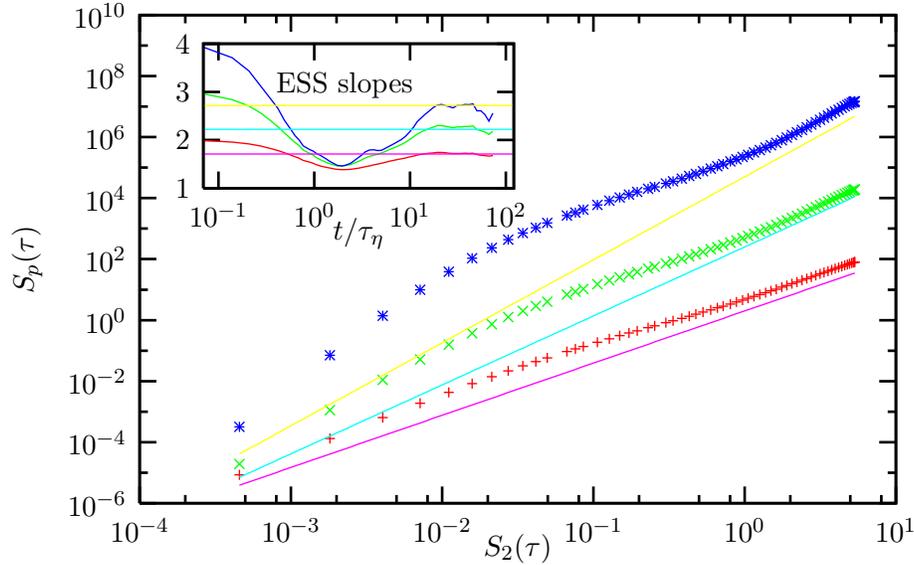}
\end{center}
\caption{ESS plot of Lagrangian velocity structure function
  $S_p(\tau)$ versus $S_2(\tau)$. Symbols refer to the DNS data for
  $p=8,6,4$ from top to bottom. Lines have slopes
  $\zeta_L(p)/\zeta_L(2)$ given by the multifractal prediction
  (\ref{eq:zeta_l}) with a $D(h)$ curve taken from the She-Leveque
  prediction \cite{she_leveque}.  In the inset, we show the local
  slopes versus time $\tau/\tau_{\eta}$, and their comparison with the
  respective multifractal prediction (straight lines). }
\label{fig:vel}
\end{figure}
%%%%%%%%%%%%%%%%%%%%%%%%%%%%%%%%%%%%%%%%%%%%%%%%%%%%%%%%%%%%%%%%
In Fig.~(\ref{fig:vel}), we present the Extended Self Similarity (ESS)
\cite{ess} log-log plot of $S_p(\tau)$ versus $S_2(\tau)$ as
calculated from our DNS. The logarithmic local slopes shown in the
inset display a deterioration of scaling quality for small times.  We
explain this strong bottleneck for time lags, $\tau \in [\tau_\eta,
10\tau_\eta]$, in terms of trapping events inside vortical structures
\cite{biferale04}: a dynamical effect which may strongly affect
scaling properties and which a simple multifractal model cannot
capture. For this reason, scaling properties are recovered only using
ESS and for large time lags, $\tau > 10 \tau_\eta$.  In this interval
a satisfactory agreement with the multifractal prediction
(\ref{eq:zeta_l}) is observed, namely from the multifractal model one
can estimate $\zeta_L(4)/\zeta_L(2) = 1.71,
\zeta_L(6)/\zeta_L(2)=2.26, \zeta_L(8)/\zeta_L(2)=2.72$ while from our DNS we
measured $\zeta_L(4)/\zeta_L(2) = 1.7\pm 0.05,
\zeta_L(6)/\zeta_L(2)=2.2 \pm 0.07, \zeta_L(8)/\zeta_L(2)=2.75\pm 0.1
$.

A similar phenomenological argument can be used to make a prediction
for the acceleration probability density function (pdf). The
acceleration can be defined as:
\begin{equation}
a \equiv \frac{\delta_{\tau_\eta} v}{\tau_\eta}.
\label{eq:a}
\end{equation}
As the Kolmogorov scale itself, $\eta$, fluctuates in the
multifractal formalism: $ \eta(h,v_0) \sim \left(\nu L_0^h/v_0
\right)^{1/(1 + h)}, $ so does the Kolmogorov time scale,
$\tau_\eta(h,v_0)$. Using (\ref{eq:tau_l}) and (\ref{eq:a}) evaluated
at $\eta$, we get for a given $h$ and $v_0$:
%%%%%%%%%%%%%%%%%%%%a%%%%%%%%%%%%%%%%%%%%%%%%%%%%%%%%%%%%%%%%%%%
\begin{figure}[!t]
\begin{center}
  \includegraphics[draft=false,scale=1.0]{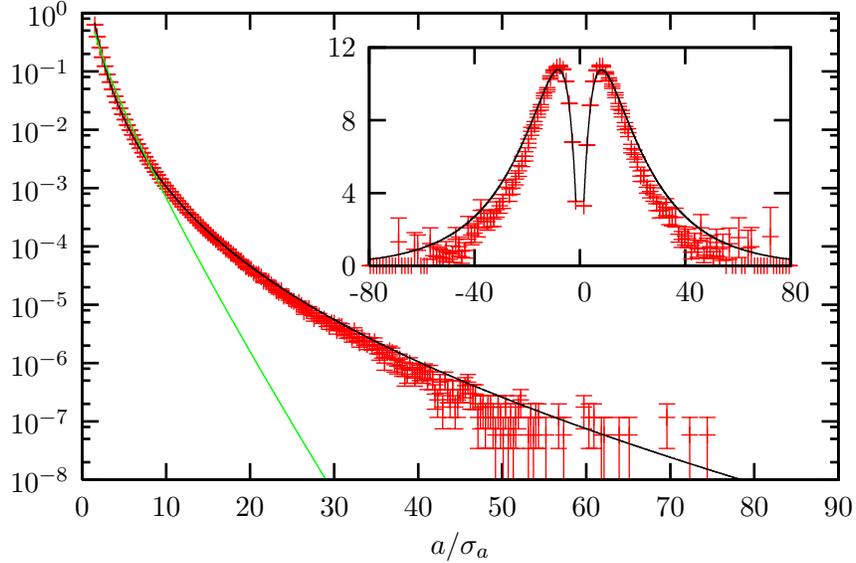}
\end{center}
\caption{
  Log-linear plot of the acceleration pdf. The crosses are the DNS
  data, the solid black line is the multifractal prediction and the
  green line is the K41 prediction. The statistical uncertainty in the
  pdf is quantified by assuming that fluctuations grow proportional to
  the square root of the number of events.  Inset: ${\tilde a}^4 {\cal
    P}({\tilde a})$ for the DNS data (crosses) and the multifractal
  prediction.}
\label{fig:acc}
\vskip -0.5cm
\end{figure}
%%%%%%%%%%%%%%%%%%%%%%%%%%%%%%%%%%%%%%%%%%%%%%%%%%%%%%%%%%%%%%%
\begin{equation}
a(h,v_0) \sim \nu^{\frac{2h - 1}{1 + h}} v_0^{\frac{3}{1 + h}} L_0^{-\frac{3h}{1 + h}}.
\label{acc}
\end{equation}
The pdf of the acceleration can be derived by integrating (\ref{acc})
over all $h$ and $v_0$, weighted with their respective probabilities,
$(\tau_{\eta}(h,v_0)/T_L(v_0))^{(3-D(h))/(1-h)}$ and ${\cal P}(v_0)$.
It remains to specify a form for the large scale velocity pdf, which
we assume to be Gaussian: ${\cal P}(v_0) = 1/\sqrt{2\pi \sigma^2_v} \,
\exp(-v_0^2/2\sigma^2_v)$, where $\sigma^2_v = \left< v_0^2 \right>$.
Integration over $v_0$ gives:
\begin{eqnarray}
&{\cal P}(a) \sim & \int_{h\in I}  \dd h \,
a^{\frac{h-5+D(h)}{3}} \nu^{\frac{7-2h-2D(h)}{3}} L_0^{D(h)+h-3} \sigma_v^{-1} \times \nonumber \\
&& \quad 
\exp \left( -\frac{a^{\frac{2(1+h)}{3}} \nu^{\frac{2(1-2h)}{3}} L_0^{2h} }{2 \sigma^2_v} \right).
\label{eq:pdf_acc}
\end{eqnarray}
In order to compare the DNS data with the multifractal prediction we
normalize the acceleration by the rms acceleration $\sigma_a=\langle
a^2\rangle^{1/2} \propto R_{\lambda}^{\chi}$. In terms of the
dimensionless acceleration, $\tilde a = a/\sigma_a$,
(\ref{eq:pdf_acc}) becomes
\begin{equation}
{\cal P}(\tilde{a}) \sim
\int _{h \in I}\tilde{a}^{\frac{(h-5+D(h))}{3}} R_{\lambda}^{y(h)}
\exp \! \left( - \frac{1}{2}\tilde{a}^{\frac{2(1+h)}{3}} 
R_{\lambda}^{z(h)} \right) \, \dd h,
\label{pdf_norm}
\end{equation}
where $y(h) = \chi (h-5+D(h))/6 + 2(2D(h)+2h-7)/3$, $z(h) = \chi
(1+h)/3 + 4(2h-1)/3$ and $\chi = \sup_h \left( 2 (D(h)-4h-1)/(1+h)
\right)$.  For more details on how the numerical integration of
(\ref{eq:pdf_acc}) is made we refer the reader to \cite{biferale04b}.

In Fig.~(\ref{fig:acc}) we compare the acceleration pdf computed from
the DNS data with the multifractal prediction (\ref{pdf_norm}). The
large number of Lagrangian particles used in the DNS ($\sim 10^6$)
allows us to detect events up to $80\sigma_a$. The accuracy of the
statistics is improved by averaging over the total duration of the
simulation and all spatial directions, since the flow is stationary
and isotropic at small-scales.  Also shown in Fig.~(\ref{fig:acc}) is
the K41 prediction for the acceleration pdf ${\cal P}_{K41}(\tilde{a})
\sim \tilde{a}^{-5/9} \exp \left( -\tilde{a}^{8/9}/2 \right)$ which
can be recovered from (\ref{pdf_norm}) with $h=1/3$, $D(h)=3$ and
$\chi=1$. As evident from Fig.~(\ref{fig:acc}), the multifractal
prediction (\ref{pdf_norm}) captures the shape of the acceleration pdf
much better than the K41 prediction. What is remarkable is that
(\ref{pdf_norm}) agrees with the DNS data well into the tails of the
distribution -- from the order of one standard deviation $\sigma_a$ up
to order $70 \sigma_a$.  This result is obtained using the
She-L\'ev\^eque model for the curve $D(h)$ \cite{she_leveque}.
%%%%%%%%%%%%%%%%%%%%%%%%%%%%%%%%%%%%%%%%%%%%%%%%%%%%%%%%%%%%%%%%
\begin{figure}[!t]
\begin{center}
  \hbox{
    \includegraphics[draft=false,width=0.50\hsize]{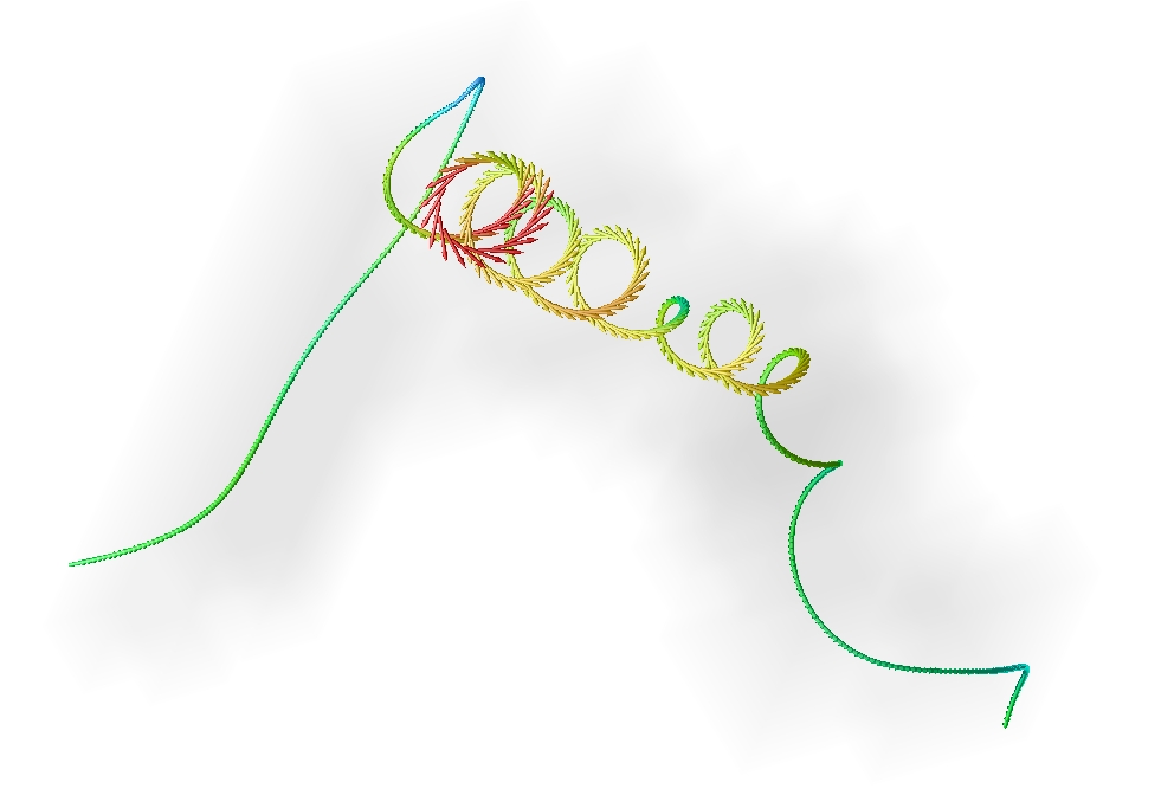}
    \includegraphics[draft=false,width=0.50\hsize]{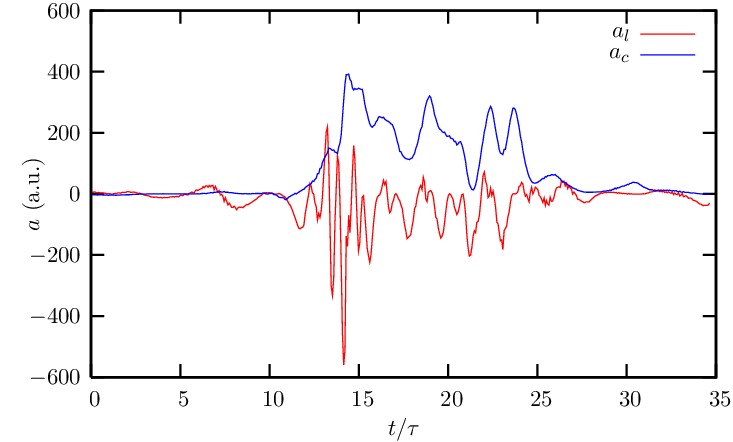}
  }
\end{center}
\caption{
  (Left panel) Trajectories with an intense value of the acceleration
  have been selected: as it can be seen, this corresponds to select
  tracers trapped into vortex filaments. Arrows and colors encode the
  velocity (magnitude and direction) of the particle.  Rendering is
  realized with OpenDX. The movie (see multimedia enhancement) shows
  the flow as seen by riding this particle, before and during the
  trapping event.  (Right panel) We show, in natural units, the
  behavior of one component of the centripetal and of the longitudinal
  acceleration (for details see the text). Notice the strong sign
  persistence of the centripetal acceleration with respect to the
  longitudinal one.}
\label{fig:traj}
\end{figure}
%%%%%%%%%%%%%%%%%%%%%%%%%%%%%%%%%%%%%%%%%%%%%%%%%%%%%%%%%%%%%%%%

\section{Acceleration tails and spiraling motion}
This and previous work \cite{cornell,pinton,biferale04} has collected
evidence which highlights the relevance to Lagrangian turbulence of
strong spiraling motions corresponding to trapping events, i.e.
passive particles trapped in small scale vortex filaments. So we
identify the strong bottleneck effect visible in Figure \ref{fig:vel}
and also the presence of extremely rare fluctuations in the pdf of the
acceleration (see Figure \ref{fig:acc}). To illustrate better these
strong events, we plot one of them in Figure \ref{fig:traj}. As is
evident, the particle while moving slowly and smoothly, at some point
gets trapped in a vortex filament and starts a spiraling motion
characterized by huge values of the acceleration and by a
``quasi-monochromatic'' signal on all the velocity field components.
Here, we suggest a way to characterize such events. This is of course
a difficult task because not all the ``trapping events'' are so
clearly detectable as that shown in Figure \ref{fig:traj}.

\begin{figure*}[!t]
\begin{center}
  \hbox{
    \includegraphics[draft=false,width=.5\hsize]{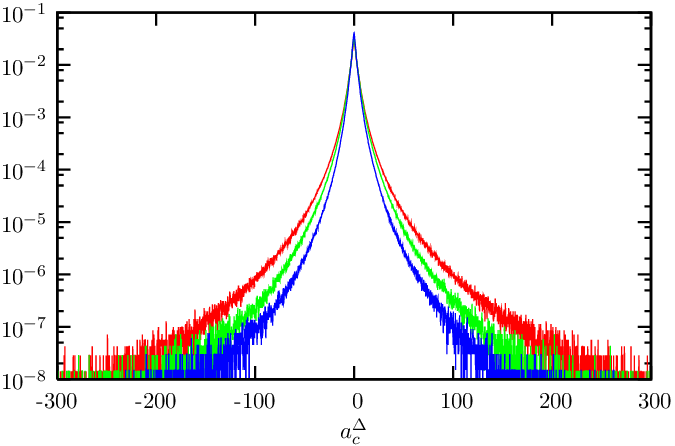}
    \includegraphics[draft=false,width=.5\hsize]{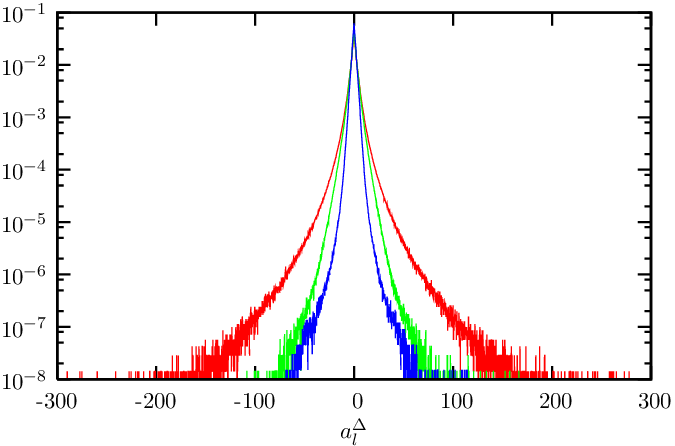}
  }
\end{center}
\caption{
  Pdf of the averaged centripetal $a_c$ (left panel), and longitudinal
  $a_l$ (right panel) acceleration components. The acceleration is
  averaged over a time window of size $\Delta = \{0.1,3,9\} \tau_{\eta}$
  (respectively corresponding to colors red, green and blue).}
\label{fig:pdf0}
\end{figure*}

Indeed the motion  of a particle in a turbulent field will be characterized
by different accelerations and decelerations, not necessarily
associated with spiraling motion (on average the mean value of the
acceleration will be zero).  In a spiraling motion the velocity ${\bm
  v}$ and acceleration ${\bm a}$ are orthogonal. Furthermore in a
circular uniform motion the angular velocity, $\omega$, can be related
to the centripetal acceleration $a_c = \omega^2 r$ and to the linear
velocity $v = \omega r$. We expect that in trapping events such as the one
depicted in Fig.  (\ref{fig:traj}) the centripetal acceleration is
intense and much more persistent than the longitudinal acceleration
(i.e. the acceleration in the direction of the motion). To make this
statement quantitative, we have studied the average of the
centripetal, ${\bm a}_c = {\bm a \times \hat {\bm v}} = {\bm a \times  {{\bm v} \over {|{\bm v}|}}}$, and
longitudinal acceleration, ${\bm a}_l = ({\bm a \cdot \hat {\bm v}}){
  \hat {\bm v}}$, over a time window which can vary up to 
$9 \tau_{\eta}$, $\Delta = \{0.1,3,9\} \tau_{\eta}$:
\begin{eqnarray}
 {\bm a}^{\Delta}_c(t) \equiv \la {\bm a}_c\ra_{\Delta} = {1\over
\Delta}\int_t^{t+\Delta} dt' {\bm a}_c(t');\\ {\bm a}^{\Delta}_l(t)
\equiv \la {\bm a}_l\ra_{\Delta}= {1\over \Delta}\int_t^{t+\Delta} dt'
{\bm a}_l(t').
\label{eq:mean}
\end{eqnarray}
We expect that the pdfs of the averaged centripetal and longitudinal
acceleration will behave very differently with increasing the window size,
$\Delta$.  In particular, the strong persistence of the
centripetal acceleration up to $10 \tau_{\eta}$ suggests that the
centripetal pdf ${\cal P}( {\bm a}^{\Delta}_c)$ should remain almost
unchanged when varying $\Delta$, while the longitudinal one ${\cal P}(
{\bm a}^{\Delta}_l)$ should become less and less intermittent.  This
is what we show in Fig.~(\ref{fig:pdf0}).
%%%%%%%%%%%%%%%%%%%%%%%%%%%%%%%%%%%%%%%%%%%%%%%%%%%%%%%%%%%%%%%

\begin{figure*}[!t]
\begin{center}
\hbox{
\includegraphics[draft=false,width=0.33\hsize]{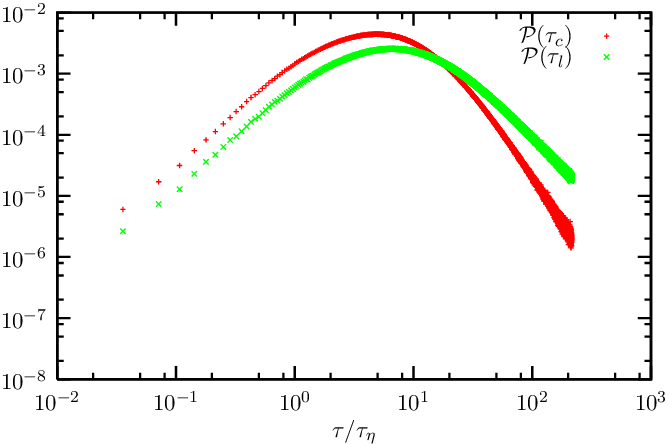}
\includegraphics[draft=false,width=0.33\hsize]{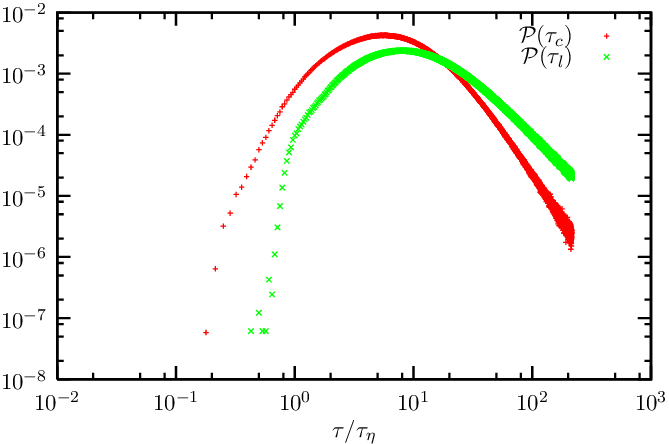}
\includegraphics[draft=false,width=0.33\hsize]{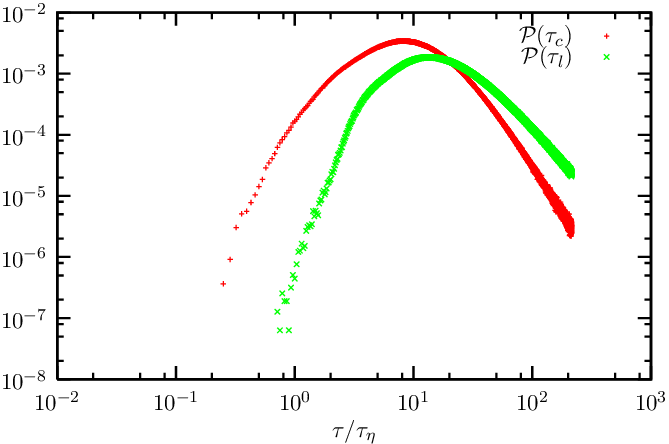}
}
\end{center}
\caption{
  Pdf of the characteristic time estimated from the centripetal (in
  red) and longitudinal (in green) accelerations (in units of
  $\tau_{\eta}$) ${\cal P}(\tau_c)$ and ${\cal P}(\tau_l)$
  respectively, for $\Delta=0.1 \tau_{\eta}$ (left panel); $\Delta=3
  \tau_{\eta}$ (central panel); $\Delta= 9 \tau_{\eta}$ (right
  panel).}
\label{fig:pdf1}
\end{figure*}

In order to investigate further the role of trapping in vortices, we
can define a typical radius of gyration $r_c$ and its typical eddy
turnover time $\tau_c$, as:
\begin{equation}
\label{eqn:gyra}
r_c ={ |{\bm v}|^2 \over {\left| {\bm a} \times {\hat {\bm v}}
\right|}}\;\;\;\;\;\;\;\mbox{and}\;\;\;\;\;\;\; \tau_c = {{\left|
{\bm v}\right|} \over {\left| {\bm a} \times {\hat {\bm v}} \right|}}
\end{equation}
Notice that using ${\bm a}\times {\hat {\bm v}}$ corresponds to
selecting the centripetal values of the acceleration and hence
augmenting the signal/noise ratio of spiraling motions with respect to
the background of turbulent motions. The previous expressions applied
to a typical vortex filament give $r_c \sim \eta$ and $\tau_c \sim
\tau_{\eta}$.  Similarly one may define a typical time based on the
``longitudinal acceleration'': $\tau_l = {{\left| {\bm v}\right|} /
  {\left| ({\bm a} \cdot {\hat {\bm v}} )\hat {\bm v} \right|}}$.
Incoherent fluctuations with typical times of the order of
$\tau_{\eta}$ should be averaged out once we measure the {\it mean}
centripetal and longitudinal accelerations averaged over a window with
$\Delta > \tau$ in expression (\ref{eqn:gyra}). On the other hand, the
signal coming from coherent vortex should not be affected by the
averaging procedure and keeps its value: as a consequence, we should
see events with $\tau_c \sim \tau_{\eta}$ even upon averaging.  Going
through Figure \ref{fig:pdf1} we can observe, with increasing window size, the different behaviors of the pdfs of the
centripetal and longitudinal characteristic times, $\tau_c$ and
$\tau_l$ respectively. It is interesting to notice that the left
tail of the centripetal pdf is quite robust, showing the presence of
characteristic times of the order of $\tau_c \sim \tau_{\eta}$ even
after averaging over a window with $\Delta = 9 \tau_{\eta}$. On the
other hand the longitudinal characteristic times of order $\tau_l \sim
\tau_{\eta}$ soon disappear as long as $\Delta \ge \tau_{\eta}$. We
interpret this as further evidence of the importance of trapping
in vortex filaments.
%%%%%%%%%%%%%%%%%%%%%%%%%%%%%%%%%%%%%%%%%%%%%%%%%%%%%%%%%%%%%%%
%%%%%%%%%%%%%%%%%%%%%%%%%%%%%%%%%%%%%%%%%%%%%%%%%%%%%%%%%%%%%%%
\section{Conclusions}
We have presented results on the Lagrangian single-particle statistics
from DNS of fully developed turbulence.  In particular we have shown
that (i) in the large time lag limit, $10 \tau_{\eta} < \tau < T_L$,
velocity structure functions are well reproduced by a standard
adaptation of the Eulerian multifractal formalism to the Lagrangian
framework; (ii) the acceleration statistics are also well captured by
the multifractal prediction; (iii) for time lags of the order of the
Kolmogorov time scale, $\tau_{\eta}$, up to time lags $10
\tau_{\eta}$, the trapping by persistent vortex filaments may strongly
affect the particle statistics. The last statement is supported both
by the scaling of the Lagrangian statistics and by a new analysis
based on the centripetal and longitudinal acceleration statistics.
\section*{Acknowledgement}
We thank the supercomputing center CINECA (Bologna, Italy) and
the ``Centro Ricerche e Studi Enrico Fermi'' for the resources
allocated for this project. We also aknowledge C. Cavazzoni, G.
Erbacci and N. Tantalo for precious technical assistance.


\section*{References}
\begin{thebibliography}{99}
\bibitem{K65} Kraichnan RH 1965 {\em Phys. Fluids} {\bf 8} 575.
\bibitem{pope} Pope SB 2000 {\it Turbulent Flows} (Cambridge
  University Press, Cambridge).
\bibitem{cornell} La Porta A, Voth GA, Crawford AM, Alexander J and
  Bodenshatz E 2001 {\em Nature} {\bf 409} 1017.  Voth GA {\it et al}
  2002 {\em J. Fluid Mech.} {\bf 469} 121.  Mordant N {\it et al.}
  2003 {\em Physica D} {\bf 193} 245.
\bibitem{pinton} Mordant N {\it et al.} 2003 {\em J. Stat. Phys.} {\bf
    113} 701. Mordant N {\it et al.} 2002 {\em Phys. Rev. Lett.}  {\bf
    89} 254502.  Mordant N {\it et al.} 2001 {\em Phys. Rev.  Lett.}
  {\bf 87} 214501.  
\bibitem{ott_mann} Ott S and Mann J 2000 {\em J. Fluid Mech.} {\bf
    422} 207.
\bibitem{leveque} Chevillard L, Roux SG, Leveque E {\it et al.}  2003
  {\em Phys. Rev. Lett.} {\bf 91} 214502.
\bibitem{yeung} Yeung PK 2002 {\em Ann. Rev. Fluid Mech.} {\bf 34}
  115.  Yeung PK 2001 {\em J. Fluid Mech.} {\bf 427} 241.  Vedula P
  and Yeung PK 1999 {\em Phys. Fluids} {\bf 11} 1208.
\bibitem{BS02} Boffetta G. and Sokolov IM 2002 {\em Phys. Rev. Lett.}
  {\bf 88} 094501.
\bibitem{IK02} Ishihara T and Kaneda Y 2002 {\em Phys. Fluids} {\bf
    14} L69.
\bibitem{beck} Beck C 2003 {\em Europhys. Lett.} {\bf 64} 151.  Beck C
  2001 {\em Phys. Lett. A} {\bf 27} 240.
\bibitem{aringazin} Aringazin AK and Mazhitov MI 2003 {\em Phys. Lett.
    A} {\bf 313} 284.
\bibitem{arimitsu} Arimitsu T and Arimitsu N 2003 {\em Physica D} {\bf
    193} 218.
\bibitem{GK03} Gotoh T and Kraichnan RH 2004 {\em Physica D} {\bf 193}
  231.
\bibitem{biferale04} Biferale L, Boffetta G, Celani A, Lanotte A and
  Toschi F 2004 \emph{Particle trapping in fully developed turbulence}
  {http://arxiv.org/abs/nlin.CD/0402032}.
\bibitem{biferale04b} Biferale L, Boffetta G, Celani A, Devenish BJ,
  Lanotte A, and Toschi F 2004 {\em Phys. Rev. Lett.} {\bf 93} 064502.
\bibitem{frisch} Frisch U 1995 {\it Turbulence: the legacy of A.N.
    Kolmogorov} (Cambridge University Press, Cambridge).
\bibitem{borgas93} Borgas MS 1993 {\em Phil. Trans. R. Soc. Lond. A}
  {\bf 342} 379.
\bibitem{BDM02} Boffetta G {\it et al.} 2002 {\em Phys. Rev. E} {\bf
    66} 066307.
\bibitem{ess} Benzi R, Ciliberto S, Tripiccione R, Baudet C, Massaioli
  F and Succi S 1993 {\it Phys. Rev. E} {\bf 48} R29
\bibitem{she_leveque} She ZS and L\'ev\^eque E 1994 {\em Phys. Rev.
    Lett.} {\bf 72} 336.
\end{thebibliography}
\end{document}